\documentclass[preprint]{aastex61}
\pdfoutput=1 
\usepackage{amsmath,amstext}
\usepackage[T1]{fontenc}
\usepackage{apjfonts} 
\usepackage[figure,figure*]{hypcap}
\usepackage{url} 
\usepackage{amsmath}
\usepackage{booktabs}
\usepackage{hyperref}
\usepackage{subfigure}
\usepackage{enumerate}
\usepackage{comment}
\usepackage{xspace}
\usepackage{soul}
\usepackage{threeparttable}
\usepackage[colorinlistoftodos]{todonotes}

\newcommand{\Rearth}{R$_\oplus$\xspace}
\newcommand{\Mearth}{M$_\oplus$\xspace}

\shorttitle{Stellar Companions \& Exoplanet Radius Distributions}
\shortauthors{Teske et al.}

\begin{document}

\title{The Effects of Stellar Companions on the Observed Transiting Exoplanet Radius Distribution}

\author{Johanna K. Teske}
\altaffiliation{Hubble Fellow, jteske@carnegiescience.edu} 
\affiliation{Department of Terrestrial Magnetism, Carnegie Institution for Science, 5241 Broad Branch Road, NW, Washington, DC 20015}
\affiliation{Observatories of the Carnegie Institution for Science, 813 Santa Barbara Street, Pasadena, CA 91101}
\author{David R. Ciardi}
\affiliation{Caltech/IPAC-NASA Exoplanet Science Institute Pasadena, CA, USA}
\author{Steve B. Howell}
\affiliation{NASA Ames Research Center, Mountain View, CA, USA}
\author{Lea A. Hirsch}
\affiliation{University of California, Berkeley, 510 Campbell Hall, Astronomy Department, Berkeley, CA 94720}
\author{Rachel A. Johnson}
\affiliation{NASA Ames Research Center, Mountain View, CA, USA}
\affiliation{University of Toledo, Toledo, OH, USA}

\begin{abstract}
Understanding the distribution and occurrence rate of small planets was a fundamental goal of the \textit{Kepler} transiting exoplanet mission, and could be improved with  \textit{K2} and TESS. Deriving accurate exoplanetary radii requires accurate measurements of the host star radii and the planetary transit depths, including accounting for any ``third light'' in the system due to nearby bound companions or background stars. High-resolution imaging of \textit{Kepler} and \textit{K2} planet candidate hosts to detect very close (within $\sim0.5$\arcsec) background or bound stellar companions has been crucial for both confirming the planetary nature of candidates, and the determination of accurate planetary radii and mean densities. Here we present an investigation of the effect of close companions, both detected and undetected, on the 
observed (raw count) exoplanet radius distribution. We demonstrate that the recently detected ``gap'' in the observed radius distribution (also seen in the completeness-corrected distribution) is fairly robust to undetected stellar companions, given that all of the systems in the sample have undergone some kind of vetting with high-resolution imaging. However, while the gap in the observed sample is not erased or shifted, it is partially filled in after accounting for possible undetected stellar companions. These findings have implications for the most likely core composition, and thus formation location, of super-Earth and sub-Neptune planets. Furthermore, we show that without high-resolution imaging of planet candidate host stars, the shape of the observed exoplanet radius distribution will be incorrectly inferred, for both \textit{Kepler}- and TESS-detected systems.  

\end{abstract}


\section{Introduction} \label{sec:intro}
\subsection{The Radius Gap}
    The \textit{Kepler} mission, the first dedicated space-based search for exoplanets, revolutionized our understanding of planet formation by detecting hundreds of super-Earth and sub-Neptune-sized planets \citep[e.g.,][]{batalha2014,thompson2017}. The \textit{Kepler} observations indicate that, for orbital periods $\lesssim$400 days, small planets (1-4~\Rearth) are much more frequent in the Galaxy than larger, Saturn and Jupiter-sized planets \citep{howard2012,dressing&charbonneau2013,fressin2013,petigura2013,ciardi2015,burke2015}, which were previously the most commonly detected planets \citep[e.g.,][]{marcy2005,udry2007,wright2012}. 
    
Furthermore, \cite{fulton2017} (hereafter F17) recently showed that by incorporating more precise, uniformly-derived stellar parameters (and thus stellar radii estimates) than the original \textit{Kepler} Input Catalog (KIC) values, a bimodality in both the observed and intrinsic distributions of small planets, previously hidden by larger planet radii uncertainties, is exposed. The authors detect a gap in the radius distribution between 1.5 and 2~\Rearth, and determine that planets above and below the gap have nearly equal completeness-corrected occurrence rates but those within the gap have an occurrence rate decreased by $\gtrsim 50$\%. The location of this gap in both the observed and intrinsic planet radius distributions is noteworthy because it occurs around the radius (1.6~\Rearth) at which planets are thought to shift from being rocky to gaseous \citep{rogers2015,marcy2014}.  

A gap in the intrinsic planetary radius distribution between 1.5 and 2.5~\Rearth was predicted by \cite{owen&wu2013} in their theoretical study of thermal contraction and hydrodynamic evaporation of volatile envelopes. Similar studies by \cite{lopez2012} and \cite{lopezfortney2013} examined the role that thermal evolution and mass loss play in individual exoplanetary systems -- Kepler-11 and Kepler-36, respectively -- and also generalized their results to predict the frequency of planets as a coupled function of orbital period and thus XUV radiation from the host star, and core composition. However, \cite{lopezfortney2013} find a less significant and also different location of the radius gap (around 2-2.5~\Rearth) as compared to \cite{owen&wu2013}, due to the differences in parameter space exploration of Lopez \& Fortney, including many different combinations of core mass and initial composition.   

After F17 published observational evidence of a clear exoplanet radius gap, a new study by \cite{owen&wu2017} provided a simple analytical model predicting that photoevaporation of volatile envelopes naturally herds planets into two groups. The first group is comprised of planets where the hydrogen/helium envelope size is less than the core size (and less than a few percent mass) of the planet and is thus stripped away, leaving a bare core. The second group is comprised of planets where the hydrogen/helium envelope is roughly the same size as the core (and a few percent mass) of the planet and the timescale for mass loss is longest. By assuming a constant, Earth-like core composition and a distribution of core sizes centered at 3~\Mearth, the Owen \& Wu model predicts two peaks in the planet radius distribution, coincident with those observed by F17. With a different core composition, the gap shifts, and with a range of core compositions, it is smeared out. The radius gap then appears to be a necessary outcome of both homogeneous core compositions of small planets and the photoevaporation of their volatile envelopes.

\subsection{The Role of Stellar Multiplicity in Exoplanet Radius Derivations}
\subsubsection{Detected Companions} 
In their analysis of the California Kepler Survey Sample \citep{petigura2017} of planet radii, F17 applied a series of filters, removing \textit{Kepler} Objects of Interest (KOIs) with orbital periods longer than 100 days, known false positives \citep{morton2011,morton2012,morton2016,kolbl2015}, impact parameters larger than 0.7, exoplanets around dim stars, exoplanets around giant stars, and planets orbiting stars with effective temperatures below 4700 K and above 6500 K. These filters resulted in a sample size decreased from 2025 KOIs with well-characterized parameters to 900 after the filtering process. The corresponding exoplanet radii were then re-derived using the light curve parameters of \cite{mullally2015}. 

In all of the \textit{Kepler} data releases (DRs) \citep{mullally2015,thompson2017,thompson2018}, all KOIs are assumed to be single unless an additional entry in the KIC appears within the pipeline aperture used for the photometry, in which case the light curve is adjusted for the excess flux of the KIC star (see more in \S2.1.1). However, we know \citep[e.g.][]{adams2012,adams2013,adams2014,horch2014,cartier2015,gilliland2015,torres2015,everett2015,barclay2015,hirsch2017} that unseen stellar companions can and have influenced the determination of transiting planet radii, and that $\sim$50\% of exoplanet host stars are in multiple star systems \citep{horch2014,furlan2017,matson2018}, similar to stars not known to host exoplanets \citep[e.g.,][]{raghavan2010,duchene&kraus2013}.

If a KOI is assumed to be a single object, then any light emitted by stellar companion(s) in the same photometric aperture can contribute to the measured flux of the primary star. If a planet transits a star with an overestimated flux, the transit depth will appear shallower, and the derived planetary radius will be underestimated. This uncertainty in the measured planetary radius is augmented further by the uncertainty around which star the planet orbits (primary or secondary), especially if the distance to the secondary star, and thus whether it is actually bound to the primary star or not, is unknown. The ratio of the true planet radius to the observed radius is 
\begin{equation}
\frac{R_{p}(true)}{R_{p}(observed)} = \frac{R_{t \star}}{R_{1 \star}} \sqrt{\frac{F_{total}}{F_t}} \equiv X_R
\end{equation}

\noindent where $R_{1 \star}$ is the radius of the primary star, and $R_{t \star}$ and $F_t$ are the radius and brightness of the star the planet is actually transiting \citep{ciardi2015}.

F17 looked into removing from their sample the KOIs with known companions or large dilution corrections, but found no significant difference in the resulting observed exoplanet radius distribution and chose not to filter their catalog based on high-resolution imaging. The compilation of high-resolution imaging the F17 authors referenced is \cite{furlan2017}, which consists of 1903 primary KOIs and their 2297 known companions observed by various sources, including the \textit{Kepler} Follow-Up Observation Program \citep{lillo-box2012,lillo-box2014,horch2012,horch2014,everett2015,gilliland2015,cartier2015,wang2015a,want2015b,kraus2016,baranec2016,ziegler2017a,ziegler2017b,ziegler2018a,adams2012,adams2013,dressing2014,law2014,howell2011,horch2011}. These observations -- mostly from near-infrared AO and optical speckle -- of the separation, magnitude difference, and position angle between primary and companion stars were used by Furlan et al. to calculate correction factors ($X_R$, as defined in Eq. 1) for planet radii taking into account the ``third light'' contamination of the stellar companions. These factors were calculated under two separate assumptions -- the planets orbit the primary star (Furlan, Table 9) and the planets orbit the detected stellar companion (Furlan, Table 10). 

The sample in the Furlan catalog represents a biased group of the ``most interesting'' targets for planet confirmation, and is not complete. However, they find that $\sim$10\% of KOIs in their sample have a stellar companion within 1$\arcsec$ and $\sim 30$\% have a companion within 4$\arcsec$ (one \textit{Kepler} pixel). The observed fraction of stellar companions is expected to be \textit{lower} than the actual fraction due to sensitivity and completeness limitations \citep{furlan2017}. That is, based on sample selection, observing conditions, and the sensitivity and resolution of the available instruments, the true fraction of KOIs with companions is expected to be higher than these fractions, especially considering companions that are faint ($\Delta$mag$\gtrsim 6-8$) and/or very close ($\lesssim 0.1 \arcsec$ projected separation) to the primary star. This expectation motivates the work described in this paper to help quantify the effects of undetected companions.

\subsubsection{Undetected Companions} 
\vspace{-5pt}

What effect, then, do \textit{undetected} companions have on exoplanet radius estimates? \cite{ciardi2015} investigated this question for gravitationally bound companions, calculating probabilistic radius correction factors for planets based on expected stellar multiplicity rates and companion parameters from studies of field stellar populations. First Ciardi et al.~identified an appropriate isochrone for each KOI in the 23 October 2014 Kepler catalog, and then considered as viable companions all of the stars following the same isochrone with absolute Kepler magnitudes fainter than the target KOI. These potential fainter companions were used to derive the planetary radius corrections considering six multiplicity scenarios: a single star ($X_R = 1$), a binary system in which the planet orbits the primary star, a binary system in which the planet orbits the companion, a triple star system in which the planet orbits the primary star, a triple system in which the planet orbits the secondary star, and a triple star system in which the planet orbits the tertiary star. In cases in which the planet orbited the primary star, only the flux dilution factor (second term in Eq. 1) was relevant, since in this case $R_{1 \star} = R_{t \star}$.

Second, \cite{ciardi2015} calculated the mean radius correction factor across the six multiplicity scenarios 
for each KOI by (1) fitting a third order polynomial to the radius correction factor versus mass ratio for each individual multiplicity scenario, (2) convolving each multiplicity scenario polynomial fit with the mass ratio distribution from \cite{raghavan2010}, (3) calculating a weighted mean for each multiplicity scenario for each KOI, and (4) convolving the six scenario corrections with the probability of the star being single (54\%), a binary (34\%), or a triple (12\%) star \citep{raghavan2010}. In multi-star systems, \cite{ciardi2015} assumed that the planet was equally likely to orbit any one of the stars. While the mean correction factor $\langle X_R \rangle$ depends on host star temperature, the authors estimate that, on average and \textit{assuming no ground-based follow-up}, the radii of KOIs are underestimated by an average factor of $\sim 1.5$ due to undetected companions. 
    
As described below, all of the KOIs in the F17 filtered observed sample have some form of ground-based follow-up observations to search for instances of ``third light'' in the \textit{Kepler} photometric aperture. These follow-up observations will decrease the number of undetected companions, and thus the predicted radius correction factors. \cite{ciardi2015} take this into account by assuming the following vetting observations: A few radial velocity (RV) observations over 6-9 months that are able to detect stellar companions with $\sim 2$ year orbital periods or less, and high-resolution imaging observations that are able to detect stellar companions with separations of $\gtrsim$0.1\arcsec. The authors then use the orbital period distribution of stellar companions from \cite{raghavan2010} combined with estimates of the distance to each KOI from the observed and absolute \textit{Kepler} magnitudes (the latter inferred from the isochrone fitting using the Dartmouth isochrones) to estimate the fraction of undetected companions for each KOI. The $X_R$ factors are then recalculated, assuming that detected companions have already been corrected for in the planet radius determination, by replacing the strict probability of a star being a multiple (46\%; from \citealt{raghavan2010} but see also \S3.3) with the probability that it is a multiple and the companion is undetected. The new $\langle X_R \rangle$, assuming the ground-based vetting observations described above, is $\sim$1.20, lower than the unvetted case but still significantly above unity.

Our study builds on the framework of \cite{ciardi2015} to examine how undetected companions might affect the distribution of raw planet counts as a function of radius (the \textit{observed or raw count} versus completeness-corrected exoplanet radius distribution presented in F17). In \S2.1.1, we first apply radius correction factors, mostly from \cite{furlan2017}, to KOIs that are in the F17 sample. These correction factors are for detected stellar companions, and we assume that the KOIs with detected companions harbor no additional undetected companions. In \S2.1.2, we take the remaining KOIs without detected companions, apply a modified version of the Ciardi et al. radius correction factors, and show how this affects the observed exoplanet radius distribution. In \S2.2, we recalculate the modified radius correction factors assuming the KOIs are at a closer distance, more akin to the likely TESS sample of planet host stars, and show how these corrections have a smaller effect on the observed exoplanet radius distribution. Finally in \S3, we discuss how the application of the radius correction factors influences the robustness of the radius gap in the observed sample (versus completeness-corrected sample) and possible small planet formation scenarios, comment on our assumptions about multiplicity of planet host vs. non-host stars, and consider the broader implications for future high-resolution imaging follow-up observations of exoplanet host stars. We summarize our results in \S4. 

\section{Methods \& Results} \label{sec:methods}

\subsection{California Kepler Survey Sample}
F17 compared their sample to the \citet{furlan2017} high-resolution imaging catalog, and found no significant change to their observed planet radius distribution by removing KOI hosts with known companions or large dilution corrections. Ultimately they chose not to filter their catalog using high-resolution imaging results. In this section, we want to answer the question, how do stellar companions affect the bimodal observed exoplanet radius distribution found in F17? (As noted in F17, it is not straightforward to fold stellar multiplicity into occurrence rate calculations, and we do not take on that task here, focusing only on the observed, ``raw counts'' exoplanet radius distribution.) For completeness we investigate the effect of both detected and undetected companions. To account for the effect of \textit{detected companions}, we cross-matched the \textit{Kepler} host star sample with high-resolution imaging observations cataloged by \citet{furlan2017} and \citet{ziegler2018b} and applied average radius correction factors calculated as in Furlan et al.. To account for the effect of \textit{undetected companions} we use a prescription modified from \citet{ciardi2015} to calculate updated radius correction factors, and applied these values. 

\subsubsection{Detected Companions}
Using an updated list of KOIs with high-resolution imaging  \citep{furlan2017,ziegler2018b}, we verify that \textit{all} of the 900 KOIs in the F17 filtered sample have some kind of high-resolution imaging follow-up, and of those 321 have detected companions within 4\arcsec. As described above, \cite{furlan2017} calculated the average exoplanet radius correction for each KOI from various imaging observations, with the average weighted by the inverse of the square of the uncertainty of each observation (each bandpass). These average radius correction factors are calculated assuming the planet orbits the primary star (their Table 9), and  assuming the planet orbits the brightest companion star (their Table 10).  

To accurately account for the stellar companions reported in \citet{furlan2017} and \cite{ziegler2018b}, we first checked whether any of these systems were ``unblended'' as defined by Furlan et al., that is, whether the detected companion star was a distinct source in the KIC. If a detected companion has a KIC value, this means that during pre-search data conditioning module (PDC) of the \textit{Kepler} data processing pipeline, the excess flux from the companion is accounted for and the light curve adjusted before it is fit for a planet radius \citep[Susan Mullally, priv. comm.]{keplermanual}. If this were the case, we would not want to account for the diluted flux a second time. 

For the 321 KOIs with detected companions in the F17 sample, only two of the companions had different KIC identifications, meaning that for these KOIs, their companions are accounted for in the reported planet radii. These stars -- KOI 1901 and KOI 4792 -- have detected companions at 3.85$\arcsec$ and 3.81\arcsec, respectively, and are removed when we applied below a constraint on the separation of companions. Thus there is no ``double'' correction applied in these cases. 


We next checked the angular separation of the detected companions, and filtered out KOIs with companions outside of a specific separation, either 1$\arcsec$ or 2\arcsec. The probability of a companion star being bound decreases as its spatial separation from the primary star increases  -- inside 0.25\arcsec, $\sim100$\% of companions are bound, inside 1\arcsec, up to 80\% of companions are likely to be bound, and within 2\arcsec, $\lesssim50$\% of companions are likely to be bound \citep{horch2014,hirsch2017,matson2018}. Whether a companion is bound or not matters for calculating the radius correction under the assumption that the planet orbits the companion star, since in this case the correction factor includes the ratio of the stellar radii of the secondary and primary stars (see Eq. 6 in \citealt{furlan2017}). 

The radius of the secondary star can be determined using multi-color observations and isochrone fitting to interpolate from the primary star's stellar parameters \citep[e.g.,][]{huber2014} to the companion's parameters \citep{everett2015,hirsch2017}. However, this analysis is only possible if the two stars are bound and assumed to fall on the same isochrone. For a background star with unknown distance and interstellar extinction, it is much more difficult to accurately assess the stellar parameters, including stellar radius. We therefore chose to apply detected companion radius correction factors from \cite{furlan2017} (and those calculated based on the updated observations of \cite{ziegler2018b}) to stars with companions within 1$\arcsec$ and 2\arcsec, respectively, as these companions are most likely to be bound. Our analysis thus does not account for background companions, which exist at all separations but especially larger ones, since the radius correction factors are more difficult or impossible to calculate for these companions. \textit{Gaia} \citep{gaia2016,gaia2018} will enable better characterization of these ($\sim 1\arcsec$) background stars and calculation of their radius correction factors, but that is outside the scope of this paper. 

Furlan et al. provide two sets of radius correction factors in their work, one that is calculated assuming the planet orbits the primary star, and another assuming the planet orbits the secondary star. In Table \ref{tab:furlan_update}, we list the average radius correction factors, under these two assumptions, for the 156 KOIs in the F17 sample that have their brightest companion within 2$\arcsec$. As the correction factor does not depend on the planet properties, we do not repeat KOIs (e.g., K00041.01 has the same $X_R$ as K00041.02 and K00041.03).

To make these radius correction factors more comparable to our modified $X_R$ values based on \cite{ciardi2015} (explained in the next section), we calculated a hybrid radius correction factor from \ref{tab:furlan_update}, $X_{R}^{primary}\times 0.7$ + $X_{R}^{secondary}\times 0.3$. In this equation, $X_{R}^{primary}$ assumes the planet orbits the primary star, and $X_{R}^{secondary}$ assumes the planet orbits the secondary star. The factors of 0.7 and 0.3 represent the probability that the planet orbits the primary versus the secondary star. Using instead an equal weighting of 0.5 and 0.5 for primary and secondary star, respectively, results in a very similar (qualitatively identical) radius distribution corrected for detected companions. We also note that an almost identical exoplanet radius distribution is produce if, instead of just using the $X_{R}^{primary}$ and $X_{R}^{secondary}$ values themselves, we draw $X_{R}^{primary}$ and $X_{R}^{secondary}$ values from a normal distribution centered on the mean $X_{R}$ values with a standard deviation equal to the uncertainties reported in \cite{furlan2017}, repeating the draw 500 times, and averaging the 500 raw count histograms together.

\begin{table*}[t]
\centering 
 \caption{Average Radius Correction Factors}
\vspace{12pt}
\footnotesize
\begin{tabular}{|c |c |c |} 
\hline
KOI & Avg Radius Correction Factor & Avg Radius Correction Factor \\
 & (planet orbits primary) & (planet orbits secondary) \\
\hline
18 & 1.0047 & 0 \\
41 & 1.0083 & 3.6043 \\
42 & 1.0349 & 1.9988 \\
72 & 1.0005 & 7.7181 \\
97 & 1.0113 & 2.2684 \\ 
105 &  1.0004 &  3.0528 \\
\hline
\end{tabular}
\tablecomments{This table is available in its entirety in a machine-readable form online. A portion is shown here for guidance regarding its form and content.}
\label{tab:furlan_update}
\end{table*}

\begin{figure}[htp]
\centering
\includegraphics[width=0.6\linewidth,clip]{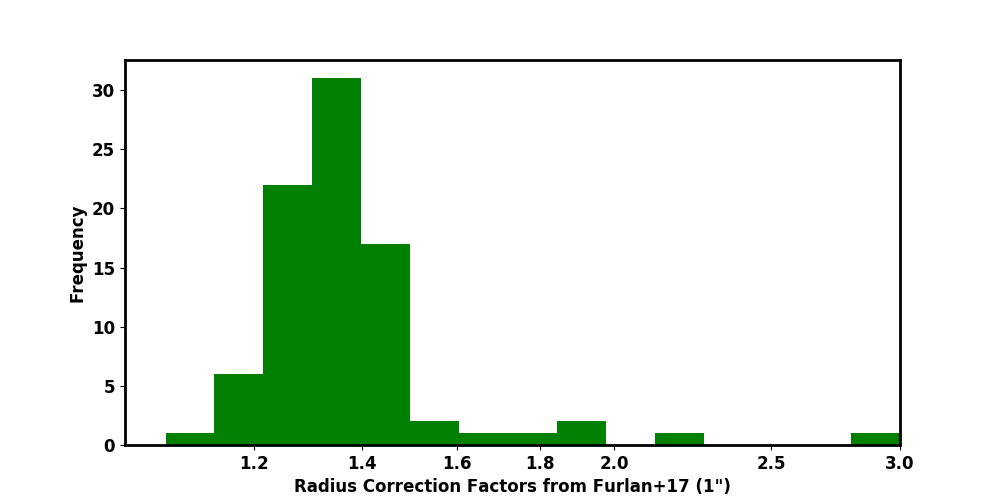}
\includegraphics[width=0.6\linewidth,clip]{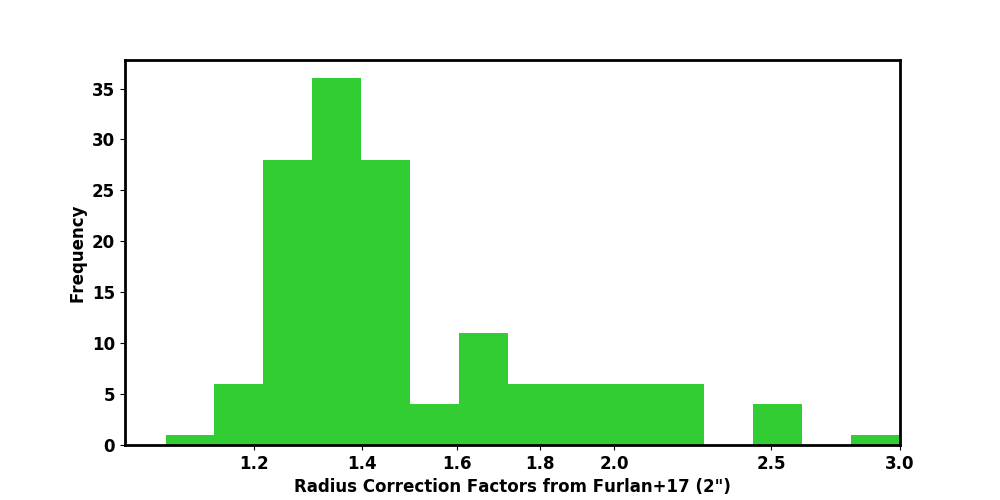}  
\caption{Distributions of the $X_R$ values corresponding to the 88 (top) or 156 (bottom) KOIs in the F17 filtered observed sample that have detected companions within 1$\arcsec$ (top) or 2$\arcsec$ (bottom).}
\label{fig:fig1}
\end{figure}

\begin{figure}[htp]
\centering
\includegraphics[width=0.7\linewidth,clip]{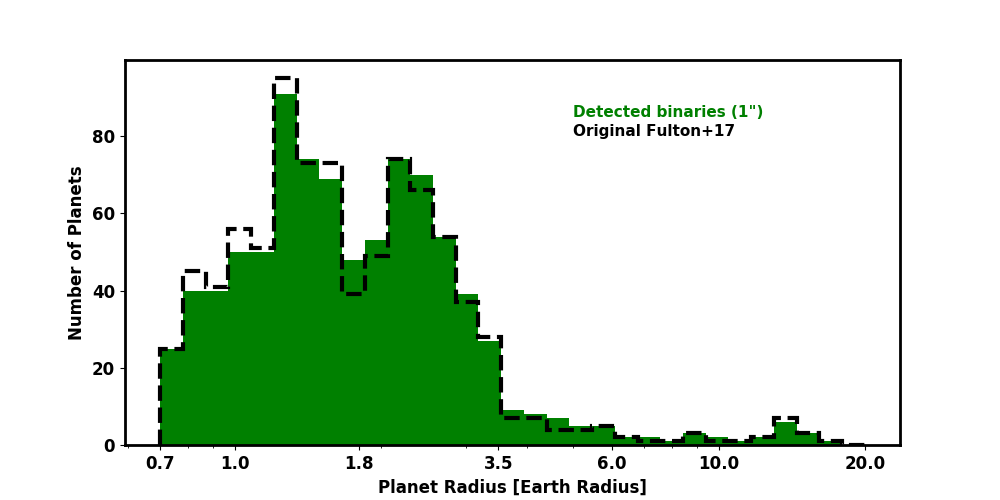}
\includegraphics[width=0.7\linewidth,clip]{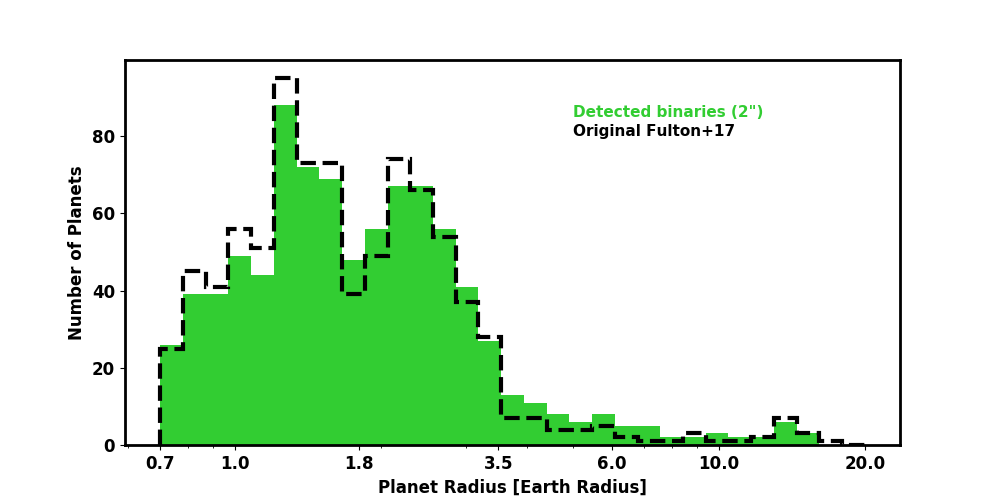}
\caption{Histograms of observed exoplanet radii from the filtered F17 sample of 900 KOIs (black dashed lines; their Figure 2 panel g), with exoplanet radius corrections applied for the detected companions from \cite{furlan2017} (filled, green histograms). Radius corrections from \citet{furlan2017} were only applied in cases where the brightest companion detected was within 1$\arcsec$ (top, 88 KOIs) or 2$\arcsec$ (bottom, 156 KOIs), and we assumed a 70/30 ratio between primary and brightest companion radius correction factors from \cite{furlan2017}. A 50/50 ratio produced similar histograms.}
\label{fig:fig2}
\end{figure}

Of the 321 KOIs in F17 with detected companions \citep{furlan2017,ziegler2018b}, 88 have the brightest companion star within 1\arcsec, and 156 have the brightest companion within 2$\arcsec$ (the rest have companions beyond 2\arcsec). In \hyperref[fig:fig1]{Figure 1} we show the distribution of $X_R$ values in the two cases (the KOI has a detected companion star within 1$\arcsec$ or within 2\arcsec), here assuming a 70/30 probability that the planet orbits the primary versus the secondary star.
In \hyperref[fig:fig2]{Figure 2} we show the resulting observed exoplanet radius distributions after applying these $X_R$ values. The colored histograms represent the corrected exoplanet radius distribution, accounting for the detected stellar companions; the original observed distribution from F17 is shown as an unfilled histogram outlined with a black dashed line. These plots show only the raw counts of planet radii, do not contain any completeness corrections, and do not represent occurrence rates. In both the 1$\arcsec$ and 2$\arcsec$ cases there is only a small change in the exoplanet radius distribution -- some $\sim$0.8-1.6~\Rearth planets shift to $\geq 1.8$~\Rearth -- and the gap does not change in position or change much in depth, as suggested by F17.

As most of the larger spatially separated ($\geq$1\arcsec) stars are unbound background (distant) sources, they tend to be fainter than the KOI and as such their brightness has little effect on the transit depth. Close ($\sim$2-4\arcsec) stars with approximately comparable brightness to the KOI will have KIC numbers and as such will have already been accounted for by the \textit{Kepler} pipeline. True bound companions, those inside 1\arcsec \citep{horch2014,hirsch2017,matson2018}, tend to be closer in brightness to the primary and therefore usually cause more significant transit dilution. Thus, moving forward, we concentrate on correcting for the detected companions within 1\arcsec.

\subsubsection{Undetected Companions}
As discussed in detail in \cite{ciardi2015}, companions around \textit{Kepler} stars can remain undetected even after vetting with high-resolution imaging and radial velocity follow-up. Below we describe how we accounted for potential undetected companions around the 812 KOIs from the filtered observed F17 sample that do not have detected companions within 1\arcsec. We assume that the 88 KOIs with detected companions within 1\arcsec, already corrected above, do not have additional undetected companions. If they did, their $X_R$ values would increase, but perhaps not significantly if the additional companion(s) had large $\Delta$magnitudes. 

The Ciardi et al. $X_R$ values were calculated under the assumptions that: (1) companions across all spectral types are equally detected, (2) each KOI could be single (their first multiplicity scenario as outlined in \S1.2.2), and (3) in the case of more than one star in the system the planet is equally likely to orbit any of the stars (50/50 in the case of a binary or 33/33/33 in the case of a triple). Then, whether or not the $X_R$ value is applied in any given case depends on the probability of the star being in a multiple system, and whether any companion stars have been detected or not. At this point, in calculating the $X_R$ values, we are interested in only the cases where the KOI is part of a multi-star system;  we do not want to include the assumption that the KOI could be single as we account for that in the next step of our method. We also adopt a different ratio $o_{prob}$ for the probability the planet orbits the primary versus a companion star. While we do not know the true $o_{prob}$ value, testing different ratios is motivated by results in the literature as well as a toy statistical logic argument, outlined below. 

As a first example from the literature, \citet{barclay2015} examined Kepler-296, a binary consisting of two M dwarfs separated by 0.2$\arcsec$ and containing five transiting planets. Using statistical and analytic arguments they found that the brighter component, Kepler-296A, is strongly preferred by the data as the exoplanet host. Kepler-13 serves as a second example -- it consists of two A-type stars, where the brighter primary (Kepler-13A) hosts a transiting planet (Kepler-13Ab), and the fainter secondary (Kepler-13B) is orbited by a third star (Kepler-13BB) of spectral type G or later \citep{shporer2014}. A substantial multi-wavelength observational effort along with detailed statistical analysis places the hot Jupiter in this system also in orbit around the primary star. Finally, \cite{fess2018} has examined 29 \textit{Kepler} multi-planet systems with high-resolution images and a detected companion, and used the transit light curves to calculate the mean density of the host star and thus assign host stars to each of the 64 planets \citep{seager2002}. Results of this study find that $\sim$90\% of the planets are statistically more likely to orbit the primary star.

Taking a back-of-the-envelope statistical approach, we find that exoplanets, especially small planets, are far more likely to be detected orbiting a brighter star versus a fainter star -- the signal-to-noise is higher and the transit depth contrast is larger around the brighter star. Dilution of the fainter star's light by the primary will also make any small planet transits around a secondary star very shallow, again reducing their chance of detection. A hard case is nearly equal brightness (mass) stars whereby any of the above techniques would not be able to differentiate between the two stars. However, in this case the planet radii will not change, regardless of which of the nearly identical stars the planet orbits.

Based on these examples and argument, we modify the original Ciardi et al. $X_R$ factors to reflect only the multi-star scenarios, and choose to test three different scenarios for the probability that the primary vs. a companion star hosts the planet, $o_{prob}$ -- 90/10, 70/30, and the original 50/50. For each $o_{prob}$ ratio, the $X_R$ mean value and spread are calculated for each KOI\footnote{We did not include the 88 KOIs with detected companions corrected in the previous section. We also did not include KOIs 163, 958, 1947, 2564, 2815, 3114, 3197, 3220, or 4457 as they were not originally in \cite{ciardi2015}.} by (1) considering all possible companions to the KOI that are fainter in absolute magnitude but could fall along the same isochrone, (2) calculating the $X_R$ factors for the possible companions assuming the planet orbits the primary, (3) calculating the $X_R$ values for the possible companions assuming the planet orbits the secondary, (4) convolving the fits of (2) and (3) vs. mass ratio with the companion-to-primary mass ratio distribution of \cite{raghavan2010} to reflect the likelihood that a companion has a particular mass and thus brightness contrast, 
and (5) taking the mean and spread of these distributions, and combining them in an average weighted by the $o_{prob}$ ratio. We choose to stay consistent with the work of \citet{ciardi2015} and use this weighted mean approach to account for the planet orbiting the primary vs. secondary star.

\begin{figure}[htp]
\centering
\includegraphics[width=0.68\linewidth,clip]{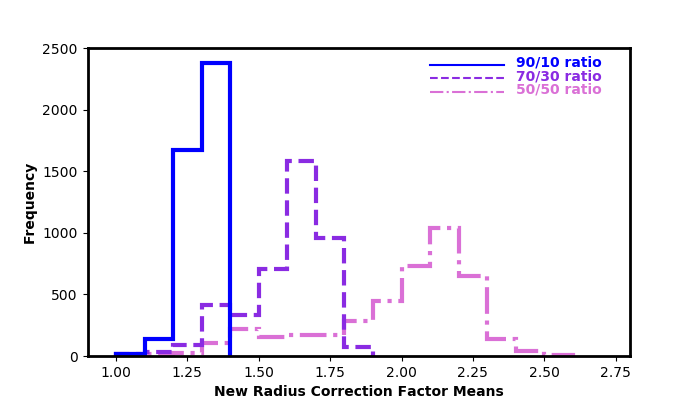}
\includegraphics[width=0.68\linewidth,clip]{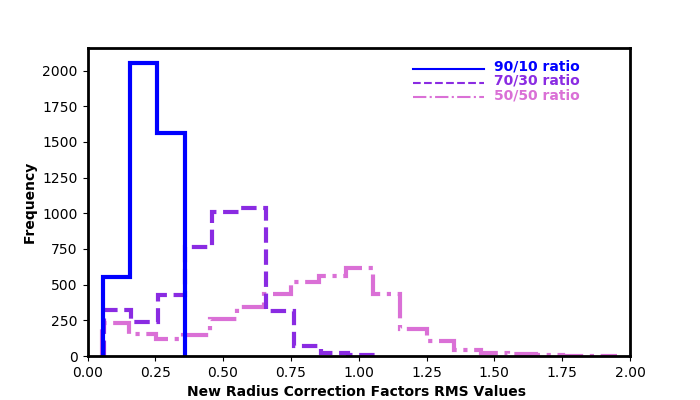}
\caption{Each new $X_R$ calculated for this study consists of a mean and a spread. In the top plot we show the distribution of mean values, in the bottom plot we show the distribution of RMS values. In blue (solid) lines are the $X_R$ values assuming a 90/10 probability ratio for the primary vs. a companion star hosting the planet, in violet (dashed) lines are the $X_R$ values assuming a 70/30 ratio, and in orchid (dash-dot) lines are the $X_R$ values assuming a 50/50 ratio.}
\label{fig:fig3}
\end{figure}

These final $X_R$ mean and spread values are listed in \hyperref[tab:modxr]{Table 1} and shown in \hyperref[fig:fig3]{Figure 3}, where the blue (solid line), violet (dashed line), and orchid (dashed dotted line) histograms correspond to $o_{prob}$ of 90/10, 70/30, and 50/50, respectively. To capture the effect of scatter in the $X_R$ values calculated from this multi-step process, for each KOI we then create a 1000-element normal distribution with the corresponding $X_R$ mean and spread, referred to as dist$_{xr}$, which is always truncated at 1 to prevent any $X_R$ values $<1$. 

\begin{table*}[t]
\centering 
 \caption{Modified Radius Correction Factors}
\vspace{12pt}
\footnotesize
\begin{tabular}{|c |c |c |c |c |c |c |c |c |} 
\hline
KOI & Fraction of multis & Fraction of multis &$X_R$ mean&$X_R$ rms&$X_R$ mean&$X_R$ rms&$X_R$ mean&$X_R$ rms\\
 & not removed& not removed & 50/50 & 50/50 & 70/30 & 70/30 & 90/10 & 90/10 \\
 & by vetting& by vetting (TESS) &  &  &  &  &  &  \\
\hline
2 & 0.290&0.054&1.459&0.438&1.312&0.287&1.204&0.175 \\
3&0.083&0.000&1.874&0.508&1.538&0.322&1.2770&0.181 \\
7&0.350&0.1026&1.989&0.957&1.584&0.600&1.278&0.259 \\
10&0.423&0.166&1.540&0.493&1.352&0.319&1.209&0.182 \\
17&0.378&0.126&1.581&0.513&1.372&0.328&1.213&0.184 \\
\hline
\end{tabular}
\tablecomments{Here we list (right-most columns) the different radius correction factors we calculated for each KOI, given different $o_{prob}$ values (50/50, 70/30, or 90/10). In the second and third columns, we list the fraction of companion stars not removed by the assumed vetted (ground-based RV and high-resolution imaging follow-up), in the case of typical \textit{Kepler} and TESS distances, respectively. This table is available in its entirety in a machine-readable form online. A portion is shown here for guidance regarding its form and content.}
\label{tab:modxr}
\end{table*}

With the $X_R$ distributions in hand, we next determine the chance that a given star is in a multi-star system and thus when we need to multiply the planet radius by $X_R$. We choose to assume (and know in the high-resolution imaging case) that all of the KOIs have been vetted with ground-based follow-up, and that any companions that \textit{could} have been detected \textit{were} detected. If a star was not vetted, the probability of it being in a multi-system can be estimated at 46\%, 
based on both field stars and the observed binary fraction of \textit{Kepler} host stars \citep{raghavan2010,horch2014,matson2018}. In the case of vetting, this number has to be multiplied by the fraction of multiple stars that have \textit{not} already been detected/accounted. We adopt the fraction of multiples not removed for each KOI from \cite{ciardi2015}, where they assumed all companions with periods of $\lesssim 2$ years and separations of $\gtrsim 0.1 \arcsec$ were detected; these values are listed in \hyperref[tab:modxr]{Table 1}, second column, ``Fraction of multis not removed by vetting''. The fraction of multis not removed by vetting is then multiplied by 0.46 to represent the remaining probability that a KOI has an undetected companion in the vetted case we are considering. We refer to this final value as prob$_{multi}$. Finally, to calculate a probabilistic $X_R$ value for each KOI, we draw a random number $r$ out of 1000. If $r\leq 1000\times$prob$_{multi}$, we then draw a random value from dist$_{xr}$, which we call $X_R^{multi}$, and multiply the exoplanet radius by this value. If $r>1000\times$prob$_{multi}$, then we assign $X_R = 1$ and do not change the exoplanet radius. 

\begin{figure}[htp]
\begin{center}
\includegraphics[width=1\linewidth,clip]{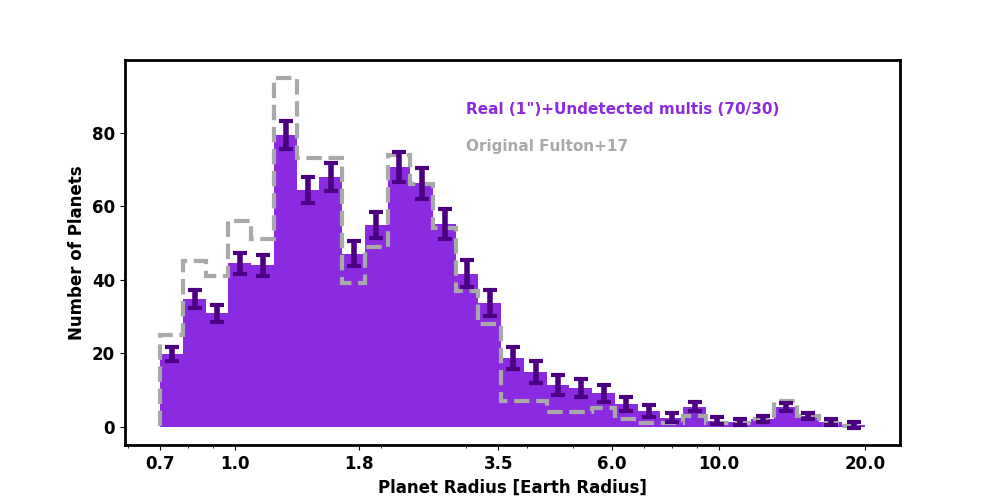}
\caption{Histogram of observed exoplanet radii from the filtered F17 sample of 900 KOIs (grey dashed line; their Figure 2 panel g), with exoplanet radius corrections applied for both detected and undetected companions (filled histogram). The height of each bar represents the average across 1000 repetitions of a different random draw of $r$, and then, if appropriate, from dist$_{xr}$, calculated assuming $o_{prob}$ = 70/30. The error bar represents the scatter in the values in each bin across the 1000 repetitions. In these plots, radius corrections from \citet{furlan2017} were only applied in cases where the brightest companion detected was within 1\arcsec, and we assumed a 70/30 weighting.}
\end{center}
\label{fig:fig5}
\end{figure}

\begin{figure}[htp]
\begin{center}
\includegraphics[width=1\linewidth,clip]{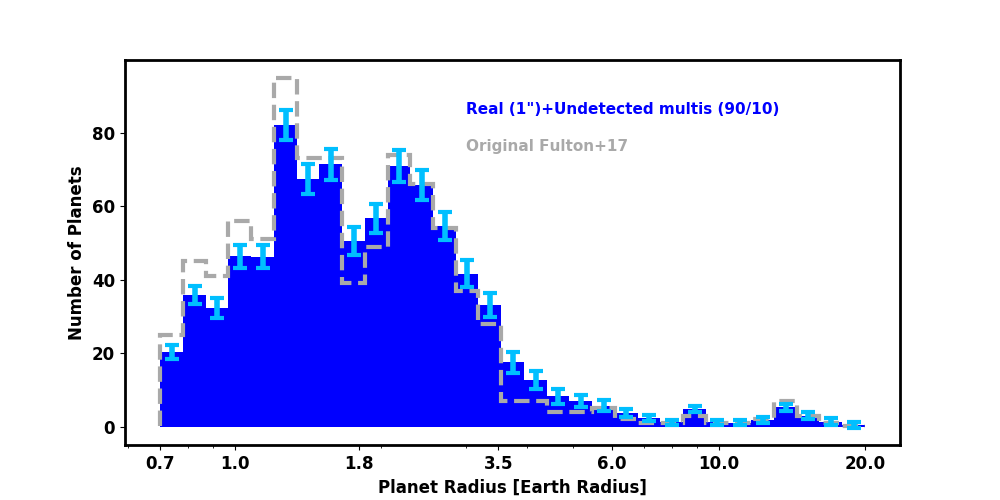}
\caption{Same as \hyperref[fig:fig5]{Figure 4}, except dist$_{xr}$ are calculated assuming $o_{prob}$ = 90/10. In these plots, radius corrections from \citet{furlan2017} were only applied in cases where the brightest companion detected was within 1\arcsec, and we assumed a 90/10 weighting.}
\end{center}
\label{fig:fig6}
\end{figure}

\begin{figure}[htp]
\begin{center}
\includegraphics[width=1\linewidth,clip]{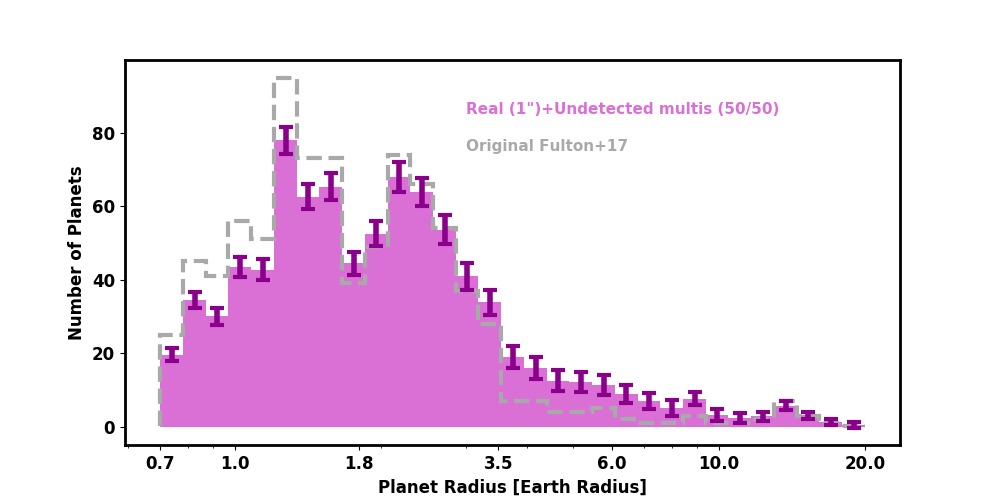}
\caption{Same as \hyperref[fig:fig5]{Figure 4}, except dist$_{xr}$ are calculated assuming $o_{prob}$ = 50/50. In these plots, radius corrections from \citet{furlan2017} were only applied in cases where the brightest companion detected was within 1\arcsec, and we assumed a 50/50 weighting.}
\end{center}
\label{fig:fig7}
\end{figure}

\begin{figure}[htp]
\begin{center}
\includegraphics[width=1\linewidth,clip]{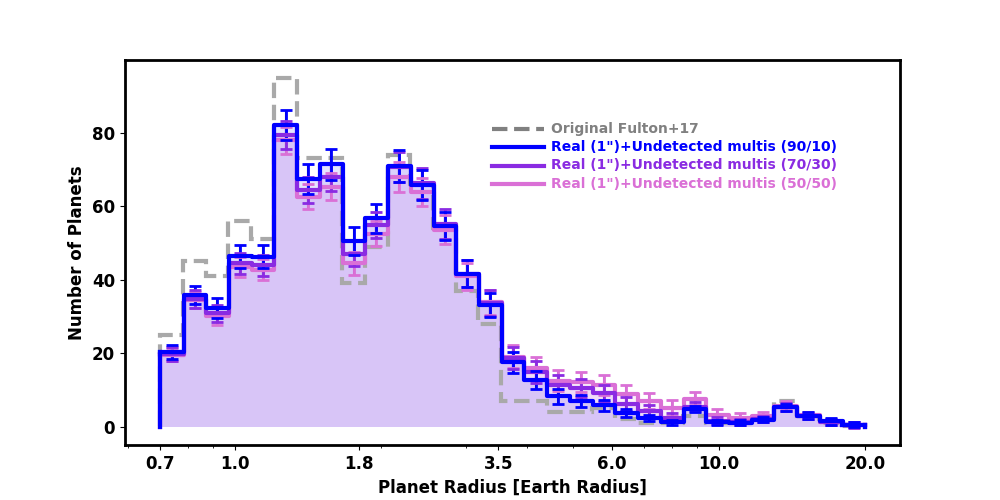}
\caption{This figure shows a comparison of the histograms in \hyperref[fig:fig5]{Figure 4}, \hyperref[fig:fig6]{Figure 5}, and \hyperref[fig:fig7]{Figure 6}, as well as the histogram of observed exoplanet radii from the F17 sample of 900 KOIs as a dark grey dashed line.}
\end{center}
\label{fig:fig8}
\end{figure}

Applying the procedure above to the 812 KOIs without detected companions within 1\arcsec results in a new histogram of exoplanet radii. To this histogram, we add back in the planet radii that were already corrected for detected companions using the $X_R$ values from \cite{furlan2017}, from \S2.1.1. We then repeat the creation of the new histogram -- accounting for both possible undetected companions and adding back in the detected companions -- 1000 times, resulting in 1000 values for each bin in the exoplanet radii histogram. The mean and spread of each bin are represented as colored histograms in \hyperref[fig:fig5]{Figure 4}, \hyperref[fig:fig6]{Figure 5}, and \hyperref[fig:fig7]{Figure 6}, each representing a different $o_{prob}$ ratio (70/30, 90/10, 50/50). The original observed distribution from F17 is shown in each figure as an unfilled histogram outlined with gray dashed line.
Note that we recalculated the $X_R$ values for the \textit{detected} companions in the previous section assuming the three different $o_{prob}$ values (different weightings of the Furlan et al. Tables 9 and 10). 
Again, these plots show only the raw counts of planet radii and do not contain any completeness corrections. In \hyperref[fig:fig8]{Figure 7} we also show all three distributions together for ease of comparison, along with the original F17 observed exoplanet distribution as a gray dashed line.  

The ``true'' effect of detected and undetected companions on the raw count exoplanet radius distribution falls somewhere within the error bars in Figures \hyperref[fig:fig5]{4}-\hyperref[fig:fig7]{6}. For each of the $o_{prob}$ values, there is some filling in of the gap, as well as a shift in the smallest planet radii to larger values, as expected since the radius correction factor is always $\geq 1$. The different $o_{prob}$ cases agree within errors, except for the $\sim$5.5-6.1~\Rearth bins where the 90/10 and 50/50 cases do not overlap within errors. Also, in the $o_{prob}$ = 50/50 case, the trend is for more of the $R_p \leq$1.7~\Rearth planets to be shifted to $R_p \geq$3.5~\Rearth, versus the $o_{prob}$ = 90/10 case, where the trend is for more of the $R_p \leq$1.6~\Rearth planets to be shifted to 1.7~\Rearth$\geq R_p \leq$2.1~\Rearth, within the gap. However, in most cases the raw count radius gap is preserved (though less distinct), as is the drop-off of planet frequency around 3.5~\Rearth (though the total frequency of planets larger than 3.5~\Rearth increases).  

\subsection{Predictions for TESS}

The recently-launched Transiting Exoplanet Survey Satellite \citep[TESS,][]{ricker2015} is focused on detecting planets around the brightest stars across the entire sky. As pointed out by \citet{ciardi2015}, because the stars will be $\sim10\times$ closer, the effectiveness of high-resolution imaging will improve greatly, decreasing the fraction of undetected companions from $\sim$40\% in the case of \textit{Kepler} to $\sim$16\% in the case of TESS. To understand how undetected companions might affect the exoplanet radius distribution observed by TESS, we can apply the same procedure as we did in the real KOI case above, calculating modified $X_R$ factors, but assuming distances 10$\times$ closer, which changes the probability that a star will have an undetected companion.  

\begin{figure}[h!]
\begin{center}
\includegraphics[width=1\linewidth,clip]{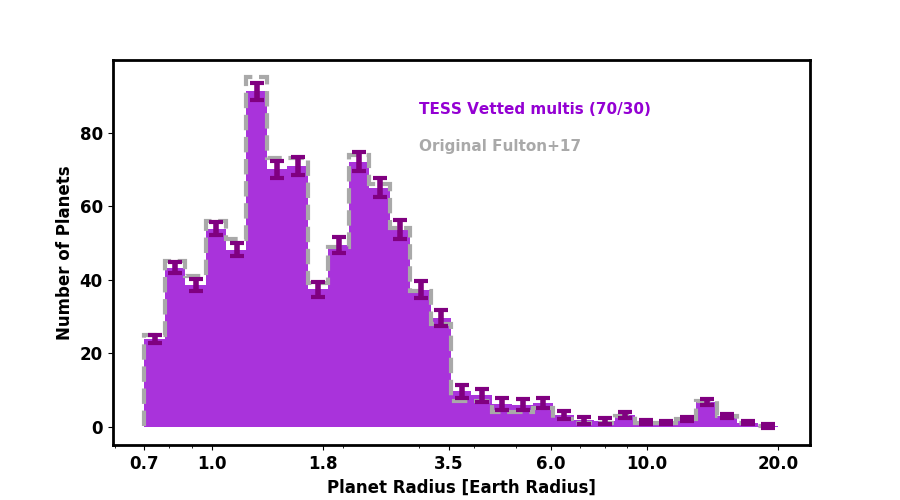}
\caption{Same as \hyperref[fig:fig5]{Figure 4}, except prob$_{multi}$ values are calculated assuming a more TESS-like target distance and the analysis is applied to all 900 KOIs (except KOIs 163, 958, 1947, 2564, 2815, 3114, 3197, 3220, or 4457 as they were not originally in \cite{ciardi2015}).}
\end{center}
\label{fig:fig9}
\end{figure}

In this case we do not have a detected companion sub-sample, so we apply the scheme outlined above to all 900 KOIs in the observed filtered F17 sample, except KOIs 163, 958, 1947, 2564, 2815, 3114, 3197, 3220, or 4457 as they were not originally in \cite{ciardi2015}. We also choose to set $o_{prob}=$70/30 for these calculations. The results are shown in \hyperref[fig:fig9]{Figure 8}; again, the raw counts of planet radii are plotted with no attempt to correct for completeness, and the F17 filtered observed sample is outlined with a gray dashed line. With high-resolution imaging follow-up that reaches well within separations of 1\arcsec, there is almost no difference between the corrected exoplanet radius distribution (colored histogram) and that not accounting for undetected companions (grey dashed line). 

\begin{figure}[h!]
\begin{center}
\includegraphics[width=1\linewidth,clip]{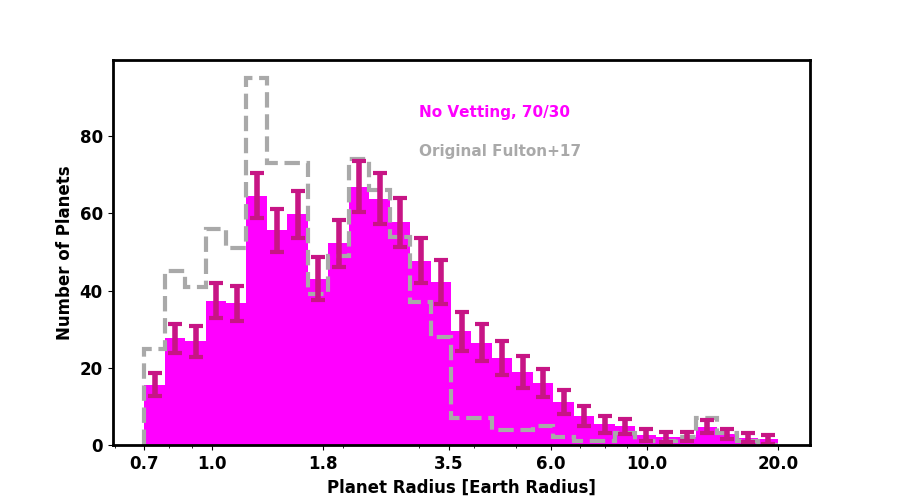}
\caption{Same as \hyperref[fig:fig5]{Figure 4} or \hyperref[fig:fig9]{Figure 8}, except prob$_{multi}$ values are calculated assuming \textit{no} vetting, that is, prob$_{multi}$ is always$=0.46$.}
\end{center}
\label{fig:fig10}
\end{figure}

What happens if there is \textit{not} high-resolution imaging follow-up of TESS targets? We investigate this scenario by assuming that prob$_{multi}$ = 0.46 for all stars; that is, none of the possible companions around a star have been detected or ruled out. We recalculate the observed exoplanet radius distribution for TESS under this assumption. The result, shown in \hyperref[fig:fig10]{Figure 9}, is that the corrected exoplanet radius distribution (colored histogram) differs significantly from the distribution inferred without accounting for undetected companions (grey dashed line). We conclude that one would infer a different and likely incorrect radius distribution if none of the companions were detected. 

\section{Discussion} \label{sec:discussion}

\subsection{Radius Gap Robustness}

We describe our method to account for detected and undetected stellar companions to KOIs in \S2.1 based on high-resolution imaging observations, comparing the isochrones of KOIs and considering viable companion stars, informed by statistics of stellar multiplicity in the field and in the \textit{Kepler} and \textit{K2} samples. This scheme, the results of which are shown in Figures \hyperref[fig:fig5]{4}-\hyperref[fig:fig8]{7}, tends to partially fill in the ``gap'' in the observed exoplanet radius distribution around 1.8~\Rearth, diluting it but not erasing or significantly shifting it. The robust nature of the observed radius gap to detected and undetected companions is likely due in part to sample selection and the vetting of that sample. F17's sample includes the ``best and brightest'' targets, those for which they were able to obtain high-resolution optical spectroscopy and successfully determine more precise stellar parameters than originally derived in the KIC (from photometry). As the F17 authors describe, their sample is also filtered for false positives, and as we confirm, all of their final sample have high-resolution imaging observations. 

We recalculate the observed exoplanet radius distribution from F17 but assume there was no vetting -- that the probability of a star being in a multiple system (prob$_{multi}$) is always 46\% -- with the results shown in Figure \hyperref[fig:fig10]{9}. (The result is the same whether the stars are a \textit{Kepler}- or TESS-like distances.) The distribution is not as easily distinguishable as bimodal, and shows a larger count of $R_p > 3$~\Rearth planets. This exercise highlights, for stars at any distance, the importance of proper vetting of planet candidate host stars with high-resolution imaging.

\subsection{Implications for Planet Formation}
It is instructive to consider what it means to ``dilute'' the radius gap, even slightly, as changing the radius gap may have implications for the average core composition of super-Earth and sub-Neptune sized planets. Numerous papers have shown that an ``evaporation valley'' in the radius distribution is a natural outcome of the photo-evaporation and thus mass loss of small planets' volatile-rich envelopes due to high-energy radiation from the host star \citep{owen&wu2017,chenrogers2016,jin2014,owen&wu2013,ciardi2013,lopezfortney2013,lopez2012}. A key parameter in these evaporation and thermal evolution models that controls the location of the valley is the core mass (or core density) of the planet, assumed not to change after formation. Since the core represents most of the mass in these planets, it controls the escape velocity and how easily an atmosphere can be evaporated. 

\cite{owen&wu2017} and \cite{jin&mordasini2017}, each using slightly different evaporation/mass loss models, found that the radius distribution of F17 was well matched by models populated with planets having uniformly rocky cores, composed of a silicate-iron mixture similar to the Earth's bulk density, and \textit{not} by planets with cores having a substantial mass fraction ($\gtrsim 75$\%) of ice/water or made purely of iron. These authors, as well as \cite{lopezfortney2013}, note that heterogeneity in the core composition would smear out the gap in the radius distribution. 

By accounting for possible undetected companions, we observe a slight smearing out of the observed radius distribution gap, particularly in the $o_{prob}$=90/10 case, which we think is the most realistic \citep{fess2018,bouma2018,barclay2015}. Our results suggests that, if there are undetected companions around the KOIs in the F17 sample, there could also be more heterogeneity in the core composition of most super-Earth and sub-Neptune planets than would be inferred from the original distribution. Specifically, a non-zero fraction of the cores could be composed of ice/water. Potential undetected companions complicate the origin story of these planets, as the addition of ice/water in the core opens up the possibility that they formed beyond the water ice line and migrated inwards, rather than only forming and migrating locally within the water ice line. Other factors not explored here, like the relative importance of X-ray/UV flux over time as a function of stellar mass, may also contribute to the radius distribution being smeared out.

We note that \cite{vaneylen2018}'s independent study of the \textit{Kepler} planet radius distribution, using a smaller sample of KOIs (75 stars, 117 planets) than F17 but with asteroseismically-derived and thus even more precise stellar parameters (and thus more precise planet radii), also finds a bimodal distribution, with two peaks at 1.5 and 2.5~\Rearth separated by a gap around 2~\Rearth, shifted to slightly higher radii than the distribution in F17. \cite{vaneylen2018} do not include any description of a correction for detected or undetected companions, but we 
 confirm that all of the KOIs in their sample have some kind of high-resolution imaging. Out of the 75 KOIs, 40 have detected companions, with separations ranging from 0.029$\arcsec$ to 3.85$\arcsec$. Of those KOIs with companions within $<1\arcsec$, the largest $X_R$ factor, assuming the planet orbits the primary star, is 1.33, and the smallest/largest $X_R$ factors, assuming the planet orbits the secondary star, are 1.38/7.02. There are seven of the 40 KOIs with detected companions that do not have enough color information to calculate $X_R$. 
It is intriguing to think that the apparent shift in the radius gap between the work of F17 and \cite{vaneylen2018} could be influenced by the effects of undetected stellar companions (although see also \citealt{fulton&petigura2018}).

\subsection{Multiplicity Assumptions}
The analysis presented here relies upon an assumption of what fraction of the planet host stars are expected to have stellar companions. We have applied radius correction factors to account for undetectable stellar companions based on the assumption that the stellar multiplicity rate of the \textit{Kepler} planet hosts is identical to the multiplicity rate for the solar neighborhood, 46\% as determined by \citet{raghavan2010}. \citet{ciardi2015} relied on the Raghavan multiplicity statistics (both the multiplicity rate and the observed distribution of companions in period and mass ratio) to simulate the \textit{Kepler} field, and the average radius correction values we apply in this work therefore depend on this assumption. 

Several studies have demonstrated that this assumption is valid, particularly for separations larger than a few tens of AU. \citet{horch2014} demonstrated that the multiplicity rate of \textit{Kepler} planet hosts as detected by the DSSI speckle camera was consistent with the solar neighborhood. With a typical resolution of 20 mas in the optical, this study was sensitive to stellar companions at separations of $\geq 20$ AU at the distance of a typical \textit{Kepler} star. \cite{matson2018} performed a similar survey for stellar companions around the somewhat nearer \textit{K2} planet hosts, and also recovered multiplicity rates similar to the solar neighborhood. We have applied radius correction factors to account for undetectable stellar companions based on the assumption that the total stellar multiplicity rate is 46\%, as determined by \citet{raghavan2010}. 

In contrast, a few surveys have reported evidence for suppressed stellar multiplicity around planet-hosting stars, for binaries with small to moderate separations. \citet{kraus2016} reports that multiplicity within 47 AU of planet hosts is suppressed by a factor of 0.34, based on a study of 382 KOIs with Keck/NIRC2 adaptive optics (AO) imaging. \citet{wang2014} also demonstrates a small suppression of stellar multiplicity at separations $\lesssim 1500$ AU, albeit for a  smaller sample of 56 KOIs but using both AO imaging and radial velocity observations. Both studies argue that this suppression of multiplicity indicates that planet formation is more difficult in close binary systems. 

If stellar multiplicity is indeed suppressed at small separations around stars hosting planets, both the average ``vetted'' radius correction factors, and the fraction of stars to which these factors are applied, would need to be altered. In other words, the number of undetectable stellar companions hiding within the CKS survey sample would be reduced. The mere observation of the gap seen by F17 may indicate that indeed the stellar multiplicity may be lower than for the general field stars. However, recent work by \cite{matson2018}, using higher resolution speckle techniques, contradict the claims of close companion suppression, showing that the fraction of close companions in \textit{K2} exoplanet host systems are similar to the field star fraction. 

\subsection{Implications for Occurrence Rate Studies}
In this work we only consider ``raw counts'', what we called the ``observed'', distribution of planet radii, and do not attempt to calculate occurrence rates. The same flux contamination effects that we described in \S1.2 for \textit{Kepler} host stars will of course also apply to stars not known to host planets within the \textit{Kepler} field, and will thus also influence the survey completeness and thus planet occurrence rates, particularly for smaller planets. While beyond the scope of this paper, we encourage future works to investigate to what extent the multiplicity of stars in the \textit{Kepler} parent sample influences the inferred planet occurrence rates. A similar high-resolution imaging survey of \textit{Kepler} non-planet hosting stars would also help determine more accurate planet occurrence rates.

\subsection{Implications for TESS Follow-up}
In the case of \textit{Kepler}, there is an orbital period/separation space in which even the best high-resolution imaging and RV follow-up do not detect companions, between $\sim$1000-100,000 days (see Figure 5 in \citealt{ciardi2015}). This is due to the \textit{Kepler} stars typically being far away, $\sim 900$ pc. Ciardi et al. calculate that, on average, ground-based observations leave $\sim 40$\% of possible companions around KOIs undetected. However, because \textit{K2} planet candidates are, and TESS Objects of Interest (TOIs) will be, closer than \textit{Kepler} targets, there is a vanishing orbital period/separation space in which high-resolution (within 1\arcsec) follow-up will not detect companions, with only $\lesssim 15$\% of stellar companions to \textit{K2} and TESS targets being missed \citep{ciardi2015,matson2018_submitted}. For comparison, the average $X_R$ calculated by \cite{ciardi2015} for vetted companions to KOIs is 1.20$\pm$0.06, while for companions to TOIs, the factor is only 1.07$\pm$0.03. We observe a similar trend in our modified $X_R$ values and the resulting exoplanet radius distributions -- with vetting there is a small but visible change in the distribution for the original KOI sample (e.g., \hyperref[fig:fig8]{Figure 7}), but there is almost no change for TESS-like distances (\hyperref[fig:fig9]{Figure 8}). 

However, as demonstrated in the \hyperref[fig:fig10]{Figure 9}, if TOIs are not vetted, the inferred exoplanet radius distribution (histogram outlined with a gray dashed line) will be different than the ``real'' distribution (filled in histograms). An incorrect distribution of planet radii will impact statistical studies of exoplanet occurrence rates and density distributions ($\rho \propto R^{3}$) \citep[e.g.,][]{furlan&howell2017}, and thus our understanding of the diversity of planets across the Galaxy. An incorrect distribution can also impact the acceptance or ``correct-ness'' of different planet formation models, as described above. 

\section{Summary}
We investigated how (bound) close companions to transiting exoplanet host stars can affect the determination of accurate planet radii, specifically the observed \textit{Kepler} small planet radius distribution with a ``gap'' around 1.8~\Rearth derived by \cite{fulton2017}. As outlined by \cite{ciardi2015}, such companions contribute to the flux measured in the photometric aperture, causing the flux of the star the planet is transiting to be overestimated, and thus the transit depth and planet radius to be underestimated. If the planet is orbiting the companion star, this can also add to the uncertainty in the inferred planet radius (see Eq. 1). The scope of this paper was limited to the study of raw planet counts, and does not include an analysis based on calculated planet occurrence rates.

First, we investigated how accounting for detected and undetected companions might change the \cite{fulton2017} observed radius distribution. We used the compilation of high-resolution observations and calculated radius correction factors from \citet{furlan2017}  and \citet{ziegler2018b} to show that correcting for \textit{detected} companions around KOIs (either 88 with companions within 1$\arcsec$ or 156 with companions within 2\arcsec) does not significantly change the observed exoplanet radius distribution (\hyperref[fig:fig2]{Figure 2}). We next modified the prescription of \cite{ciardi2015} to estimate exoplanet radius correction factors for \textit{undetected} companions, assuming (1) a multiplicity rate similar to both nearby field stars and \textit{Kepler} and \textit{K2} host stars; (2) that the KOIs were uniformly vetted for companions with orbital periods $\lesssim$2 years with RV observations and separations $\gtrsim 0.1 \arcsec$ with high-resolution imaging observations; and (3) different probabilities for the planet orbiting the primary versus secondary star ($o_{prob}$, 90/10, 70/30, or 50/50). We also assumed that the KOIs with detected companions did not have additional undetected companions. The resulting observed exoplanet radius distributions (\hyperref[fig:fig5]{Figure 4}, \hyperref[fig:fig6]{Figure 5}, and \hyperref[fig:fig7]{Figure 6}) still show the gap, but it appears to be partially filled in by the shifting of the smallest planets to larger radii (as expected, since by definition the radius correction factors are always $\geq 1$). The shape of this observed radius distribution has implications for the inferred formation pathways of small planets -- ``filling in'' the gap may indicate a more heterogeneous core composition, perhaps with some planets having water/ice material accreted from outside the snowline \citep[e.g.,][]{owen&wu2017,jin&mordasini2017}. 

Second, we applied the same undetected companion prescription to all 900 of the KOIs\footnote{except KOIs 163, 958, 1947, 2564, 2815, 3114, 3197, 3220, or 4457 as they were not originally in \cite{ciardi2015}} in the Fulton et al. observed filtered sample, but assumed a distance 10$\times$ closer, more similar to the stars that TESS will survey for transiting planets. We show that with high-resolution imaging vetting ($\lesssim 0.25$\arcsec), there is little to no uncertainty in the observed exoplanet radius distribution (\hyperref[fig:fig9]{Figure 8}). However, without any vetting, the exoplanet radius distribution, no matter the distance to the host stars, does not match the corrected distribution (\hyperref[fig:fig10]{Figure 9}). Thus it is critical that dedicated ground-based, high-resolution imaging observations of planet candidate systems continue in the TESS era.

\acknowledgments
Support for this work was provided by NASA through Hubble Fellowship grant HST-HF2-51399.001 awarded by the Space Telescope Science Institute, which is operated by the Association of Universities for Research in Astronomy, Inc., for NASA, under contract NAS5-26555. This research has made use of the NASA Exoplanet Archive, which is operated by the California Institute of Technology, under contract with the National Aeronautics and Space Administration under the Exoplanet Exploration Program. This research has made use of the Exoplanet Follow-up Observation Program website, which is operated by the California Institute of Technology, under contract with the National Aeronautics and Space Administration under the Exoplanet Exploration Program. High-resolution images referenced in this work have been obtained at the WIYN, Gemini, Palomar, and Keck telescopes. We thank Elise Furlan for fruitful discussion and sharing information that enhanced the quality of this work, and Susan Mullally for clarifying how the \textit{Kepler} PDC accounts for stellar companions in the calculation of planet radii. We thank the anonymous referee for their detailed comments that improved the quality of this paper.  

\software{astropy}

\bibliographystyle{yahapj}
\bibliography{references}


\end{document}